\documentclass[aip,reprint]{revtex4-1}
\usepackage{amssymb}
\usepackage{graphicx}
\usepackage{caption2}

\begin{document}

\title{The Dynamics of Carbon Nanostructures at Detonation of Condensed High Explosives
}%

\author{Nataliya P. Satonkina}
 \protect\thanks{Electronic mail: snp@hydro.nsc.ru.}
\affiliation{Lavrentyev Institute of Hydrodynamics, 630090 Novosibirsk, 630090 Russia\\
Novosibirsk State University, Novosibirsk, 630090 Russia}%

\date{\today}

\newcommand\revtex{REV\TeX}

\begin{abstract}
Based on the obtained earlier experimental conductivity graphs at detonation of five different high
explosives, an analysis was performed which shows a correlation between the carbon content
and the conductivity magnitude. An assumption was put forward that the conductivity in the whole
detonation wave is provided by the contact mechanism along conductive carbon nanostructures.
\end{abstract}


\maketitle

Detonation products (DP) of condensed high explosives (HEs) of the kind C$_a$H$_b$N$_c$O$_d$ contain carbon
structures of different types: soot, diamond, amorphous carbon. The ultrafine diamond (UFD) is of
particular interest  \cite{bib:lyamkin88,bib:philips88} due to its application in industry.
Presently, there are several questions which are still not clear. Thus, there is no common opinion
about the time of nanodiamond formation, and the formation mechanism is not
clear\cite{bib:danilenko051}.

The process of carbon condensation in a detonation wave leads to the UFD formation, and this
process which is in turn connected with the kinetics of chemical reactions. Experimental
investigation of the kinetics is complicated due to intrinsic features of fast processes, namely
short duration (several microseconds), aggressiveness of the investigated medium (high pressures of
tens of GPa, high temperatures of several thousand degrees). Therefore, the kinetics is mainly
studied by the numerical methods, and its features are recovered from the final results by
investigating the structures extracted from DP after a chemical treatment.

Dynamics of the UFD formation can be traced by the electric conductivity which is the result of the
presence of carbon nanostructures as will be discussed below. Comparison of the literature data on
the chemical peak duration and the time dependence of conductivity $\sigma(t)$ has earlier shown
that if a pronounced conductivity peak is present, its duration is close to the duration of the
zone of chemical reaction \cite{bib:ershov04,bib:ershov07}. This allows one to observe the kinetics
related to the carbon almost directly based on the conductivity graph $\sigma(t)$.

Despite more than half-century research history, the nature of high conductivity at the detonation
of condensed HEs is still uncertain. There is no generally acknowledged assumption of the
predictive power. The investigation of conductivity is however highly promising. With the detailed
understanding of the mechanism of conductivity, it could become a highly sensitive and simple tool
to investigate the nanostructural changes of the medium with several advantages: only weak
disturbance of the process investigated, the diagnostics directly in the high-pressure region, and
in real time.

B. Hayes proposed in 1965 the correlation between the maximum value of the conductivity at the
detonation and the free carbon content in the DP \cite{bib:hayes65}. He also proposed the contact
mechanism of the conductivity along the carbon nanostructures which serve as ``wires''. The density
of condensed carbon was obtained numerically taking into account the compression of medium in the
detonation wave. The density was taken in the Chapman -- Jouguet point (CJP), and the maximum value
of the conductivity was used although the question about the correlation between the conductivity
and the detonation wave is still opened.

The assumption of the present paper differs from the above mentioned ones in the following point:
for the first time, the maximum value of conductivity is connected with the total carbon content,
and the conductivity in the CJP is connected with the free condensed carbon. The correlation
obtained from the experimental data allows us to claim that the conductivity in the whole
detonation wave is provided by the carbon nets in conductive phase except the cases with low mass
fraction of carbon (less than 0.06). Thus, the dynamics of conductivity tracks the evolution of
carbon nanostructures. The analysis of the experimental results obtained in the Lavrentyev
Institute of Hydrodynamics SB RAS
\cite{bib:ershov04,bib:ershov07,bib:ershov09,bib:ershov00,bib:ershov10} was carried out.


The detonation wave consists of the shock front, the adjacent chemical peak (von Neumann peak)
where chemical reactions occur, and the Taylor rarefaction wave separated from the chemical peak by
the Chapman  -- Jouguet point where the velocity of products is equal to the local speed of sound.
In the Zeldovich--von Neumann--Doering, theory, the chemical reactions are completed in the CJP.

 \begin{figure}
 \includegraphics[width=85mm]{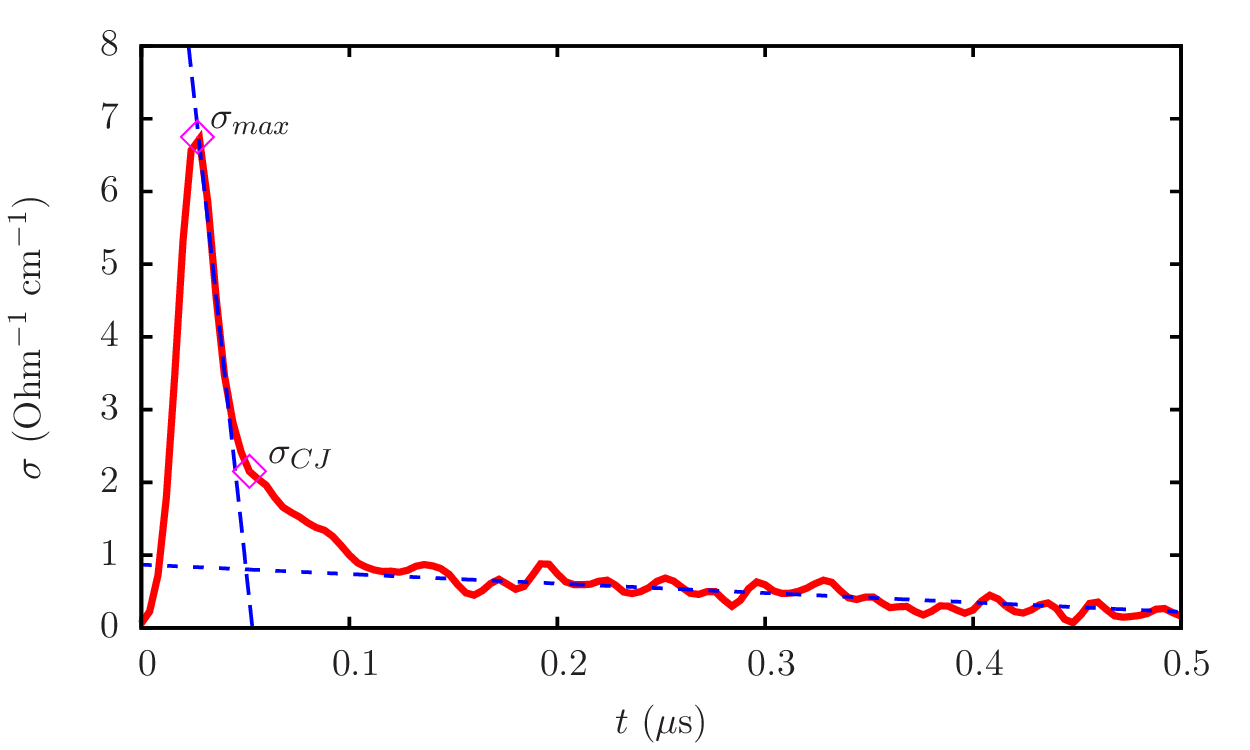}%
 \caption{Profile of electrical conductivity at  detonation of HMX, $\sigma_{max}$ - maximum conductivity, $\sigma_{CJ}$ -- electrical conductivity in CJP. The figure shows the end of the chemical
reaction zone defined by a crossing of straight lines. \label{pict1}}%
\end{figure}

The typical graph of conductivity at the detonation of condensed HE is shown in Fig. 1. The value
of conductivity increases up to the maximum $\sigma_{max}$ during several tens of nanoseconds, then
it rapidly decreases to the point marked as $\sigma_{CJ}$, which corresponds to CJP, and the region
of slowly varying conductivity in the Taylor wave. The figure shows the end of the chemical
reaction zone defined by a crossing of straight lines.

At present, there is no theoretical foundation of the connection between the conductivity graph and
the detonation wave, therefore we discuss in detail their correlation in the framework of the
assumption proposed. For the investigated HEs, the conductivity of the chemical peak $\sigma_{max}$
is greater than $\sigma_{CJ}$ \cite{bib:ershov07}. Therefore, the assumption was proposed that the
maximum conductivity $\sigma_{max}$ is provided by the total carbon content, whereas the
conductivity in the CJP $\sigma_{CJ}$ is due to the condensed carbon. Thus, the carbon structures
grow until the upper point of the chemical peak, and the reactions with carbon occur later, in the
range between the points marked as $\sigma_{max}$ and $\sigma_{CJ}$ in Fig. 1. In the CJ point,
reactions are completed, the carbon nets are thinned and partially broken, and the conductivity
$\sigma_{CJ}<\sigma_{max}$ is provided by the remaining structures. The following decrease of the
conductivity is due to the partial disruption of conducting branches in the dense DP medium and the
possible partial transition of the carbon to a non-conductive phase \cite{bib:danilenko051}.

The lower is the initial HE density, the higher is the influence of the charge inhomogeneity. In
order to connect correctly the relative fraction of carbon in the molecule with the conductivity,
the value of $\sigma_{max}$ is necessary which is hard to obtain with the experimental technique
used. Therefore we constructed approximated based on the experimental data and made an
extrapolation to the crystal density.

 \begin{figure}
 \includegraphics[width=85mm]{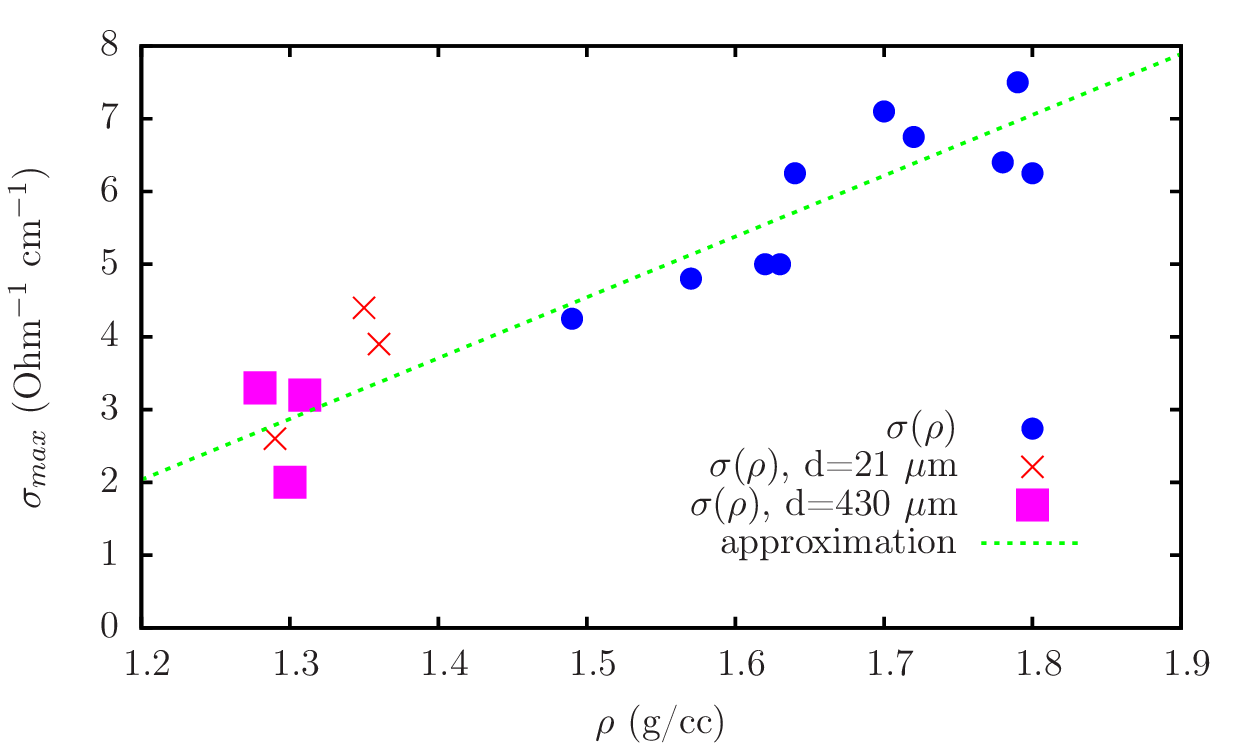}%
 \caption{The values of maximum conductivity at different initial density for HMX. \label{pict2}}%
\end{figure}

Figure 2 shows the values of $\sigma_{max}$ for HMX at different initial density and the linear
approximation. At powder densities $\rho < 1.4$ g/cc the large scatter is due to different grain
size (21 and 430 $\mu$m, \cite{bib:ershov10}) which affects both the conductivity value and the width of the
reaction zone. The increase of $\sigma_{max}$ is clearly seen. This trend is common for all the
investigated HEs.

Based on the data of \cite{bib:ershov07} where the conductivity for different densities was
obtained, we constructed the approximation of $\sigma_{max}$ for RDX, HMX and PETN. Results are
shown in Fig. 3. The maximum conductivity $\sigma_{max}$ increases with the increase of density,
which is connected with the increase of the density of carbon and carbon nanostructures.

 \begin{figure}
 \includegraphics[width=85mm]{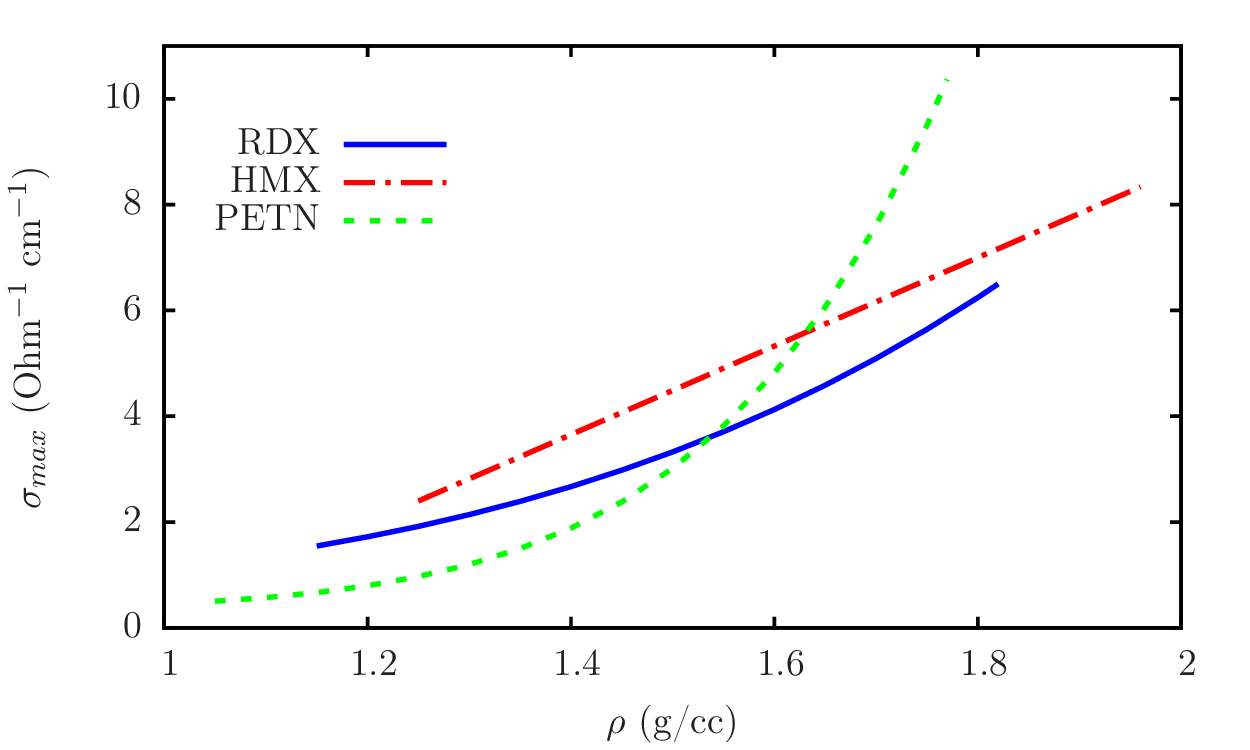}%
 \caption{The approximation of experimental data  of $\sigma_{max}$ for different HEs.\label{pict3}}%
\end{figure}

Linear extrapolation of $\sigma_{max}$ to the crystal density for TNT and TATB was made based on
the powder density and the maximum obtained one \cite{bib:ershov09,bib:safonov,bib:satonkina16}. For TNT, the results of \cite{bib:hayes65} were also taken into account.


 \begin{figure}
 \includegraphics[width=85mm]{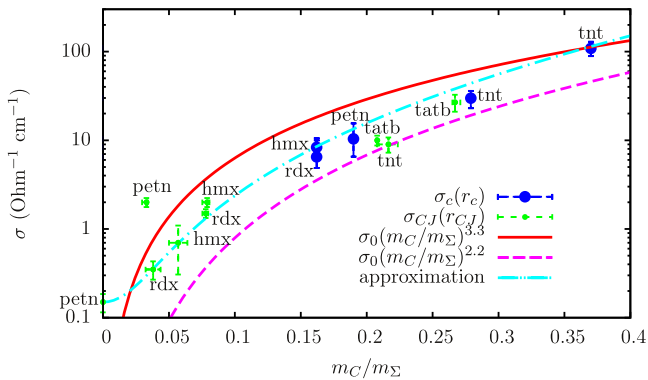}%
 \caption{ $\sigma_{c}(r_ {c})$ -- the values of maximum of conductivity at crystal density (the result of extrapolation), $\sigma_ {CJ}(r_ {CJ})$ -- electrical conductivity in CJP. The conductivity data are bounded from above  and from below.  \label{pict4}}
\end{figure}

Based on the data of \cite{bib:tanaka}, the mass fraction of carbon in the CJ point $r_{CJ}$ at
different initial density was calculated by the interpolation. Figure 4 shows some data of \cite{bib:ershov04,bib:ershov07,bib:ershov09,bib:ershov00,bib:ershov10,bib:safonov}
on the dependence of $\sigma_{CJ}$ on the mass fraction of the condensed carbon
$\sigma_{CJ}(r_{CJ})$ and the values of maximum of conductivity at crystal density $\sigma_{c}(r_ {c})$  (extrapolation results).

There is a pronounced dependence of both the maximum conductivity $\sigma_{c}$ and the conductivity
in the CJ point $\sigma_{CJ}$ on the carbon fraction. Graphs $\sigma_{c}(r_{c})$ and
$\sigma_{CJ}(r_{CJ})$ are close.

The conductivity data are bounded from above by the power-law function
$f_1=\sigma_0(m_C/m_{\Sigma})^{2.2}$ and from below by the function
$f_2=\sigma_0(m_C/m_{\Sigma})^{3.3}$. Here, $\sigma_0=1370$~Ohm$^{-1}$~cm$^{-1}$ is the conductivity
of liquid carbon. Analytical approximation of the experimental data results in a steeper dependence
and for the $m/m_{\Sigma}=1$ approaches the value of $10^4$~Ohm$^{-1}$~cm$^{-1}$. The conductivity
of highly-oriented graphite can reach the value of $2 \cdot 10^4$~Ohm$^{-1}$~cm$^{-1}$ \cite{},
hence, the extrapolation looks reasonable.

At a certain carbon fraction, the formation of connected nets become impossible due to the lack of
conductive substance. In the works \cite{bib:satonkina14,bib:satonkina141}, the threshold carbon
fraction of 0.06 was determined in numerical experiments which theoretically allows the formation
of connected nets in DP. Based of the data shown, it is possible to say that carbon influences the
conductivity also at fractions lower than 0.06. In such cases, different mechanism related to the
carbon can exist, e.g., thermal electron emission. For PETN, high value of the conductivity
$\sigma_{CJ}$ at the carbon fraction of $r_{CJ}=0.033$ could be explained by peculiarities of
chemical reactions, where the oxidation mainly leads to the thinning of structures rather than
their breakup.

The model proposed is confirmed by the following works.

1. In \cite{bib:breusov} it was shown that the formation of UFD cannot be connected with the
intermediate production of the free carbon, and the mechanism for the UFD formation was proposed
which is connected with the partial breakup of molecular bonds and the formation of the growth of
the carbon net.

2. The author of \cite{bib:anisichkin94,bib:anisichkin07} claims based on the analysis of
experiments with isotopic tag that the carbon oxidation occurs later than the formation of carbon
particles, and confirms the fast nucleation of carbon atoms. This agrees well with the data shown.

3. HEs considered in this paper have different parameters of detonation \cite{bib:tanaka}, the
detonation diamonds were found in products of most of them. When the diamond forms in the chemical
peak, this should be seen on the conductivity graph because diamond is a dielectrics
\cite{bib:elprop}. However, the dependence $\sigma_{c}(r_{c})$ has no explicit traces of the
diamond formation. Therefore, we can suppose that carbon exists in the region of the chemical peak
in a conductive phase for all the HEs investigated. The same is claimed in the work
\cite{bib:danilenko051} where it was obtained for TNT, RDX, TNT/RDX mixture, and TATB that the
conditions in the CJ point are shifted to the region of liquid nanocarbon when the size of
nanodiamond is taken into account. The author claims that diamond is formed from the liquid phase
in the rarefaction wave. In the work \cite{bib:korobenko99}, the conductivity of liquid carbon was
measured experimentally at the temperatures characteristic for the detonation. Its value of
$\sigma\approx 10^3$~Ohm$^{-1}$~cm$^{-1}$ explains the high values of conductivity $\sigma_{c}$ and
$\sigma_{CJ}$ obtained in experiments. Such value can be reached at the contact conduction along
the carbon structures connecting the interelectrode gap.

4. In the works \cite{bib:saxs}, the data on the small-angle x-ray scattering (SAXS) at
the detonation were obtained using synchrotron radiation. Since the scattering occurs on
inhomogeneities, the magnitude of SAXS is proportional to the density difference squared of the
medium and the inhomogeneity. The increase of the integral SAXS intensity is smooth without jumps
in the chemical peak region which should be present if the UFD formation is completed in the von
Neumann peak. The density of the liquid carbon phase at the temperatures of 5000 -- 7000 K and the
pressure of o.1 kbar is $\approx1.8$~g/cc \cite{bib:korobenko03}. Therefore, if the carbon
structures are liquid, the contrast is low, and the structures are not resolved due to small
density difference of DP and carbon inhomogeneities. The smooth growth can be interpreted as the
observation of the phase transition dynamics which result is the UFD.

All the mentioned confirms the model proposed.

We investigated the influence of different chemical elements on the maximum conductivity
$\sigma_{max}$ and the conductivity in the CJ point $\sigma_{CJ}$. Only the correlation with carbon
was obtained.

Analysis of the data on mixed HEs published in \cite{bib:ershov00,bib:ershov09,bib:satonkinapruuel11,bib:satonkina15} is more complicated although the overall
trend of the conductivity increase with the carbon fraction. The grain size plays a significant
role of the HEs mixed.

The problem of conductivity is multiparametric. There is a correlation with the pressure for HMX,
RDX and PETN mentioned in the work \cite{bib:ershov07}. The density and the temperature also affect
the conductivity. However, in the coordinates ``mass fraction of carbon -- conductivity'' their
role falls out, and the role of carbon can be singled out.

In the present work, the analysis of the experimental results on condensed HEs in a broad range of
the carbon fraction from 0 (in the CJ point for powder PETN) to 0.37 (fraction of carbon in TNT
molecule) was performed. We obtained a correlation between the carbon content and the conductivity
in the whole detonation wave. High values of the conductivity are explained by the contact
mechanism which is provided by conductive carbon connected nanostructures.

This work was supported by the Russian Foundation for Basic Research (project no. 15-03-01039a).


\begin{thebibliography}{9}

\bibitem{bib:lyamkin88}
A. I. Lyamkin, E. A. Petrov, A. P. Ershov, G. V. Sakovich, A. M. Staver, V. M. Titov, Sov. Phys.-Dokl. {\bf 33}, 705 (1988) [in Russian].

\bibitem{bib:philips88}
N. R. Creiner,  D. S. Philips,  J. D. Jonhnson and F. Volk, Nature, {\bf 333}, 440 (1988). 

\bibitem{bib:danilenko051}
V. V. Danilenko, Combust., Expl., Shock Waves. {\bf 41}, 577 (2005).

\bibitem{bib:ershov04}
Ershov A.P., Satonkina N.P., Ivanov G.M.
Technical Physics Letters. {\bf 30}, 1048 (2004). 

\bibitem{bib:ershov07}
 A.P. Ershov,  N.P. Satonkina, G. M. Ivanov. Rus. J. of Phys. Chem. B. {\bf 1}, 588 (2007). 

\bibitem{bib:hayes65}
B. Hayes  Proc. 4th Symp. on Det. White Oak, ACR-126, 595 (1965).

\bibitem{bib:ershov09}
A. P. Ershov, N. P. Satonkina. Combust., Expl., Shock Waves {\bf 45}, 205 (2009).

\bibitem{bib:ershov00}
A. P. Ershov, N. P. Satonkina, 0. A. Dibirov, S. V. Tsykin, and Yu. V. Yanilkin. Combust., Expl., Shock Waves.  {\bf 36}, 639 (2000).

\bibitem{bib:ershov10}
 A. P. Ershov,  N. P. Satonkina, Comb. and Flame. {\bf 157}, 1022 (2010).

\bibitem{bib:safonov}
N. P. Satonkina, A. A. Safonov, J. of Eng. Thermophysics. {\bf 18}, 177 (2009).

\bibitem{bib:satonkina16}
N. P. Satonkina, I. A. Rubtsov, Techn. Phys. {\bf 86}, in the press  (2016).

\bibitem{bib:tanaka}
K. Tanaka, Detonation Properties of Condensed Explosives Computed Using the Kihara-Hikita-Tanaka Equation of State. National Chemical Laboratory for Industry, Tsukuba Research Center (1983).

\bibitem{bib:satonkina14}
N. P. Satonkina,  A. P. Ershov, E. R. Pruuel,  D. I. Karpov, Proc.  XXIX  Int. Conf. Physics of Extreme States of Matter (2014).

\bibitem{bib:satonkina141}
N. P. Satonkina,  E. R. Pruuel,  D. I. Karpov, Proc.  XV  Int. Det. Symp. (2014).

\bibitem{bib:breusov}
O. N. Breusov, Khim. Fiz. {\bf 21  (11)}, 110 (2002) [in Russian].

\bibitem{bib:anisichkin94}
V.F. Anisichkin. Combust., Expl., Shock Waves. {\bf 30(5)}, 667 (1994).

\bibitem{bib:anisichkin07}
V. F. Anisichkin.  Combust., Expl., Shock Waves. {\bf 43}, 580 (2007).

\bibitem{bib:elprop}
A. Chaudhary, J. O. Welch, and R. B. Jackman. Appl. Phys. Lett. {\bf 96}, 242903 (2010).

\bibitem{bib:korobenko99}
 V. N. Korobenko,  A. I. Savvatimskiy and  R. Cheret,  Int. J. of Thermophysics. {\bf 20}, 1247 (1999).

\bibitem{bib:saxs}
 K. A. Ten, V. M. Aulchenko, L. A. Lukjanchikov, E. R. Pruuel, L. I. Shekhtman, B. P. Tolochko, I. L. Zhogin, V. V. Zhulanov, Nuclear Instr. and Meth. in Phys. Research A. {\bf 603}, 102 (2009).

\bibitem{bib:korobenko03}
 V. N. Korobenko,  A. I. Savvatimskiy, Temperature: Its Measurement and Control in Science and Industry. New  York. {\bf 7}, 783 (2003).

\bibitem{bib:satonkinapruuel11}
N. P. Satonkina, E. R. Pruuel, A. P. Ershov, D. I. Karpov, V. V. Sil'vestrov, A. V. Plastinin, P. A. Savrovskii, J. of Engin. Thermophysics. {\bf 20}, 315 (2011).

\bibitem{bib:satonkina15}
N. P. Satonkina, E. R. Pruuel, A. P. Ershov, V. V. Sil'vestrov, D. I. Karpov, A. V. Plastinin,  Combust., Expl., Shock Waves. {\bf 51(3)}, 1 (2015).
\end{thebibliography}
\end{document}